# In situ studies of a molten metal anode ablation in a nearly atmospheric pressure DC arc

Stanislav Musikhin[1], Valerian Nemchinsky[1], Hengfei Gu[2], Bruce E. Koel[2], and Yevgeny Raitses[1]

[1]Princeton Plasma Physics Laboratory, Princeton, New Jersey 08543, United States
[2]Department of Chemical & Biological Engineering, Princeton University, Princeton, New Jersey 08540, United States

## Abstract

A DC arc with a meltable metal anode in a near-atmospheric pressure hydrocarbon gas is an emerging method for producing single-walled carbon nanotubes (SWCNTs). In these systems, evaporation of the molten metal anode determines the formation of catalyst seed particles needed for SWCNT growth, and therefore, should be monitored, controlled, and optimized. Evaluating the anode ablation rate by weighing the anode before and after a synthesis run is unfeasible due to anode carburization in the hydrocarbon atmosphere. To overcome this, we implemented a high-speed, 2D, 2-color pyrometry for reliable temperature measurements of the molten anode in a DC arc. The obtained temperature fields were used to calculate the anode ablation rates. Results showed the importance of resolving the arc and molten pool dynamics, as well as addressing the issue of reflections. Furthermore, significant changes in ablation rates were revealed upon addition of $CH_4$, which must be considered when scaling up the production of SWCNTs.

## 1. Introduction

A nearly atmospheric or atmospheric pressure DC arc discharge is one of the main methods for large-scale production of metal and metal oxide nanoparticles, core-shell particles, and carbon nanostructures, e.g., carbon nanotubes and graphene flakes [1–5]. A DC arc offers multiple favorable features, such as design simplicity, process continuity and scalability, the ability to operate at moderate (10—100 Torr) to atmospheric pressure, no requirements for toxic precursors and growth substrates, and the ability to achieve high product chemical purity and required morphology. However, optimization of product selectivity remains a key challenge in DC arcs, requiring careful control of synthesis conditions, e.g., anode ablation. In an anodic DC arc, the anode ablation, which directly correlates with the anode surface temperature [6,7], provides the material for the growth of nanostructures and, consequently, determines their production rate and morphology. Therefore, knowledge of the anode spatial and temporal temperature evolution is crucial for monitoring, controlling, and optimizing nanomaterials synthesis. Furthermore, accurate electrode temperature measurements are required in plasma torches and thrusters, when optimizing the process to reduce electrode erosion and extend its lifetime.

Recently, DC arcs with meltable metal electrodes have been introduced for the production of single-walled carbon nanotubes (SWCNTs) [5,8]. Compared to more traditional carbon arcs utilizing a DC discharge with a graphite anode [1,2], in DC arcs with meltable metal electrodes the supply of carbon feedstock is achieved by the decomposition of a hydrocarbon gas, e.g., methane ($CH_4$), while ablation of the metal anode provides the catalyst source, e.g., iron. Such a configuration allows easy resupply of the anode material, e.g., as metal pellets or scrap, to prevent anode consumption and extend runtime [9]. However, evaluating the anode ablation rate in such systems by weighing the anode before and after a synthesis run is unfeasible due to anode carburization in the hydrocarbon atmosphere. Therefore, herein, we focus on ablation studies of the molten steel anode via accurate in situ temperature measurements, which are essential for optimal catalyst and SWCNT synthesis. Previously, we studied the growth of catalyst iron nanoparticles in a hydrocarbon environment for such arcs [5,10].

Reliable anode temperatures would also be valuable for validating electrode ablation models. Existing self-sustaining models aim to solve the electrode heat balance to determine the anode temperature, which requires minimal experimental input but involves assumptions on arc symmetry, steady-state behavior, and uncertainties in material properties at high temperatures [6,11–13]. High-speed and accurate anode temperature imaging would help to verify these assumptions, leading to more credible ablation predictions.

Several factors complicate accurate temperature measurements of molten metal electrodes, particularly in DC arcs [14]: i) High temperatures and harsh plasma environments make contact measurements, e.g., using a thermocouple, impractical; ii) Arc instabilities, such as the movement of an arc attachment spot on the anode, lead to rapid electrode temperature fluctuations that can substantially impact the morphology of synthesized nanomaterials [15]; iii) The spectral emissivity of electrode materials can be unknown or vary significantly during the measurements due to changing temperature, chemical composition, and surface roughness, e.g., the surface of either solid or molten metal, which makes single-color pyrometry measurements unreliable; iv) Plasma and particle emissions can interfere with optical measurements; and v) Reflections of surrounding hot objects from the electrode surface also lead to erroneous optical measurements.

To address these challenges, we utilized an in situ, high-speed, 2D, 2-color pyrometry setup for electrode temperature measurements. Such systems have been used in various plasmas for thermal imaging of electrodes, weld pools, and materials facing plasma in fusion devices [16–21]. The 2-color detection minimizes the error from unknown or changing material emissivity. High-speed measurements allow resolving temperature fluctuations caused by high-frequency arc instabilities. The in situ, 2D measurements enable real-time and spatially resolved measurements of the electrode temperature distribution. To avoid reflections of nearby hot objects, e.g., another electrode, optical observations can be done at an angle [17]. Although a viable solution, it is not practical in cases of limited optical access into a reaction chamber, or if a moving plasma plume contributes to reflections. To tackle these issues, we introduced a high-speed shutter, which was quickly swung in and out of a position between the electrodes to block the reflections.

This paper is organized as follows. Section 2 describes the DC arc with a meltable steel anode. Section 3 explains the pyrometry principle and presents the developed pyrometry setup design, calibration, and validation. Section 4 explains how the anode ablation rate was calculated using the obtained anode temperatures. The Results and discussion section presents temperature fields for graphite and steel anodes measured at different arc power values in Ar and in Ar/$CH_4$ gas mixtures, discusses the importance of resolving the arc and molten metal pool dynamics, and compares calculated and measured anode ablation rates.

## 2. DC arc setup

The arc setup is shown in Figure 1 and was described in detail in our previous work [5]. Briefly, the configuration consists of two vertically oriented electrodes inside a vacuum chamber. The top electrode was a 2%-ceriated tungsten cathode with a base diameter of 6.4 mm and conically shaped to ensure cathode arc spot stability. The anode at the bottom was a 9.5-mm diameter rod made of A36 steel (Fe > 99 wt.%, C 0.06 wt.%). The gap between the electrodes was monitored during the arc operation and adjusted by a stepper motor as needed.

Two gas mixtures were tested: i) Ar (5.0 purity) with a flowrate of 916 ± 5 sccm; and ii) a mixture of Ar (916 ± 5 sccm) with $CH_4$ (22.9 ± 0.1 sccm, 4.0 purity) flows, which corresponds to 2.4 wt.% $CH_4$ in Ar. Gas flows were controlled by calibrated mass flow controllers (Alicat Scientific). In both cases, the chamber pressure was monitored with a pressure transducer (MKS, Baratron 221A) and maintained at 66.7 ± 0.3 kPa (500 ± 2 Torr) using a PID pressure controller (Alicat Scientific).

The arc was ignited by a spark from a tungsten wire biased to 2 kV using a Bertan Associates Inc. (205A-05R) high-voltage power supply. The discharge was sustained with a Sorenson SGA100X100C-1AAA 100 V/100 A power supply, set to a constant current regime. Depending on the current and arc behavior, the voltage across the electrodes ranged from 10—12 V in Ar and 12—16 V in the Ar/$CH_4$ mixture. After turning off the plasma, the reactor chamber was evacuated, repressurized with Ar, and allowed to cool before opening to air in order to minimize oxidation of the electrodes.

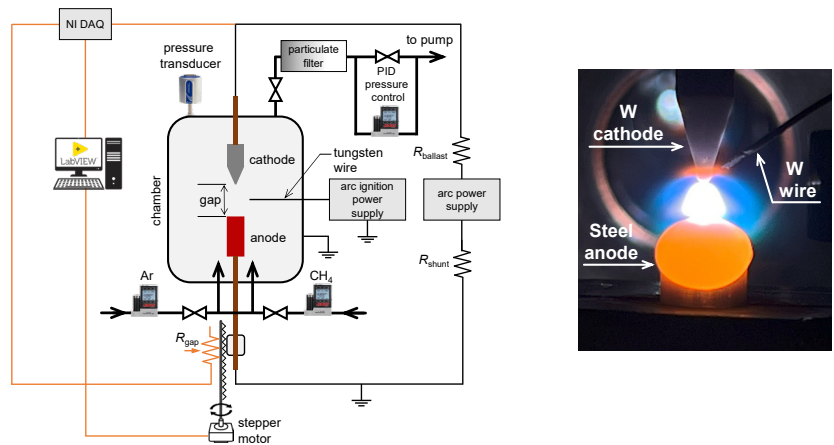

Figure 1. (Left) Diagram of the experimental setup. (Right) Photo of an arc in operation. The position of the steel anode was adjusted with a stepper motor to achieve a constant interelectrode gap as the anode melts during an experiment. The pressure inside the chamber was controlled with a PID pressure controller installed downstream of the arc discharge reaction zone.

## 3. Pyrometry setup

A schematic of the high-speed, 2D, 2-color pyrometry setup is shown in Figure 2. The design was based on ref. [22] and specifically took into account avoiding common problems of 2-color pyrometry systems. In particular, this configuration ensures both light beams travel the same optical path, eliminates the parallax effect, and only requires a single camera. Radiation from the hot electrode passes through a 50/50 beam splitter and bandpass filters, reflects off the mirrors, and finally reflects off a knife-edge prism (all optical elements from Thorlabs Inc.) that directs two parallel beams to an objective lens (Navitar, 50-mm diameter), which is mounted to a high-speed camera (Phantom v7.3, 14 bits depth). The objective lens was set to an f-number of f/11 to reduce vignetting, meaning that only the central portion of the lens was utilized. The total image size was 512 × 256 pixels with a pixel resolution of 106 µm/pixel.

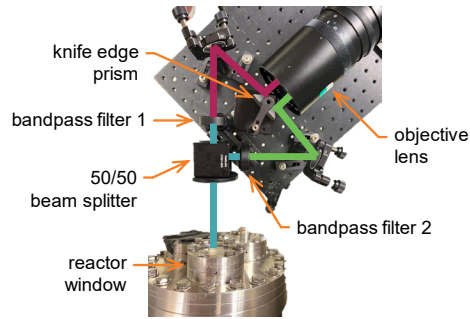

Figure 2. Optical configuration of the high-speed, 2D, 2-color pyrometry setup.

Two-color pyrometry is based on detecting thermal radiation emitted by an object at two distinct wavelengths and determining the temperature from the ratio of the radiance signals [23]. The signal intensity at a wavelength $\lambda$ is given by Wien's approximation to Planck's law:

$$I_\lambda(T) = \frac{K \varepsilon_{\lambda,T}}{\lambda^5 \exp\left(\frac{hc}{k_B \lambda T}\right)}, \qquad (1)$$

where $K$ is a system-specific constant, $\varepsilon_{\lambda,T}$ is the emissivity of an object at a wavelength λ and temperature $T$, $h$ is Planck's constant, $c$ is the speed of light in vacuum, and $k_B$ is Boltzmann's constant. For $\lambda T$ < 3000 µm K, which is true for the conditions considered herein, the error from the approximation is less than 1%. When collecting signals at two distinct wavelengths, $I_{\lambda 1}(T)$ and $I_{\lambda 2}(T)$, the temperature can be derived as:

$$T = \frac{hc(\lambda_1 - \lambda_2)}{k_B \lambda_1 \lambda_2} \frac{1}{\ln\left(I_1 \lambda_1^5 \varepsilon_2 / I_2 \lambda_2^5 \varepsilon_1\right)}, \tag{2}$$

where $I_1$ and $I_2$ are the calibrated signal intensities at the corresponding wavelengths, and $\varepsilon_1$ and $\varepsilon_2$ are the object emissivities at these wavelengths.

The choice of detection wavelengths is crucial in 2-color pyrometry. When the emissivity of the object is unknown or changes during the experiment, it is beneficial to use narrowly separated detection wavelengths, e.g., within 50 nm or closer, so that the emissivity uncertainty is minimized [24]. This is especially relevant when a grey-body assumption $\left(\frac{\varepsilon_1}{\varepsilon_2} = 1\right)$ is applied to a non-grey body. On the other hand, narrowly separated wavelengths produce radiance signals with similar intensities, making the signal ratio, and hence the temperature, highly susceptible to measurement errors. We opted to use bandpass filters at 785 ± 3 nm and 890 ± 10 nm (both from Thorlabs Inc.), also avoiding spectral interferences from atomic and molecular emission, e.g., Ar, Fe, and $C_2$ (Figure S2). The emissivity ratio for solid and liquid iron was estimated to be $\frac{\varepsilon_{785\,nm}}{\varepsilon_{890\,nm}} = 1.06 \pm 0.02$ [23,25,26]. Details on the pyrometry setup calibration, validation, and the temperature measurement uncertainty are given in the Supplementary Information.

During plasma operation, the molten anode surface acts as a mirror and specularly reflects incident radiation from the surroundings. Here, the problematic reflections were from hot radiating objects, such as the cathode, ignition wire, and dense plasma plume, which were found to corrupt the temperature measurements as demonstrated in the Results section. To block reflections and enable anode temperature measurements, a high-speed shutter was deployed. The motion system was initially developed for swinging a Langmuir probe between the electrodes [27], but was repurposed for the shutter. The system comprised a thin rectangular steel plate (shutter) mounted on a rotary feedthrough, which was connected to a DYNAMIXEL AX-18A servo motor. The plate was swung in and out between the electrodes, with an approximate residence time of 100 ms inside the plasma.

## 4. Ablation rate calculation

We observed no material ejection from the anode surface in any of the experiments, hence, the anode ablation should be explained by vaporization only. To calculate the net anode ablation rate, $\Gamma$ (kg/s), we employed the Hertz-Knudsen expression for the ablation flux [6,28],

$$g_{\text{abl}} = \sqrt{\frac{m}{2\pi k T_{\text{anode}}}} \left(P_{\text{sat}}(T_{\text{anode}}) - P_{\text{vapor}}\right), \tag{3}$$

where $g_{abl}$ denotes the net evaporation (here, ablation) flux (kg/m²/s), $m$ is the mass of the ablated species, $k$ is the Boltzmann constant, $T_{anode}$ is the measured anode surface temperature, $P_{sat}(T_{anode})$ is the saturation vapor pressure at the anode surface temperature, and $P_{vapor}$ is the vapor pressure at the Knudsen layer boundary. Many different forms of the Hertz-Knudsen equation exist in the literature. Our choice and involved assumptions are explained in the Supplementary Information. The anode ablation rate, $\Gamma$ (kg/s), is then found as, $\Gamma = A g_{abl}$, where $A$ is the ablation surface area. The calculated ablation rate was compared with the actual anode mass loss rate measured by weighing the anode before and after the experiment.

The Hertz-Knudsen relation (Eq. 3) defines the ablation flux at a liquid-vapor or solid-vapor interface. In this work, given that the measured mass loss rates were much smaller than the maximum possible ablation into a vacuum (Langmuir's case of ablation), we considered a diffusion-limited ablation, where transient effects and gas convection are negligible [29]. Therefore, the ablation flux is defined solely by vapor diffusion to the surroundings, and the following is true at the liquid-vapor or solid-vapor interface: $\Gamma_{abl} = \Gamma_{diff}$ and $g_{abl} = g_{diff}$. This ablation regime has also been called "slow evaporation" and the "low ablation regime" in the literature [6,30].

The diffusion problem is described by Fick's second law, $\frac{\partial N_{vap}}{\partial t} = D \nabla^2 N_{vap}$, where $N_{vap}$ is the vapor concentration, and $D$ is the diffusion coefficient, which was assumed to be constant here. For a steady-state case, $\frac{\partial N_{vap}}{\partial t} = 0$, the problem reduces to the Laplace equation, $D \nabla^2 N_{vap} = 0$. For symmetrical spherical diffusion, the Laplacian is given by $\frac{1}{r^2} \frac{\partial}{\partial r}\left(r^2 \frac{\partial N_{vap}}{\partial r}\right) = 0$, which has the following solution: $N_{vap}(r) = \frac{A}{r} + B$. We specify two boundary conditions: (i) the vapor concentration at the ablating surface, which is taken as a hemisphere with a radius $R$, defined as $N_{vap}(R) = N_{sat}$, and (ii) zero vapor concentration at a distance $d$, $N_{vap}(R + d) = 0$. The parameter $d$, the "condensation distance", which is sometimes called "a point of no return", is the distance the evaporated material diffuses into the surroundings before condensing [7]. Then, $A = N_{sat} \frac{R(R+d)}{d}$ and $B = -N_{sat} \frac{R}{d}$. The diffusion flux density (molecules/m²/s) then is $J_{diff} = -D \frac{\partial N_{vap}}{\partial r}\bigg|_{r=R} = D N_{sat} \frac{R+d}{Rd}$. In the case of $d \ll R$, i.e., the semi-infinite or planar case, we obtain $J_{diff} = D \frac{N_{sat}}{d}$. Correspondingly, the diffusion fluxes (kg/m²/s) are:

$$g_{diff} = mD \frac{R+d}{Rd} N_{sat} = mD \frac{R+d}{Rd} \frac{P_{sat}(T_{anode})}{k T_{anode}} \qquad (4)$$

for the spherical case, and

$$g_{diff} = \frac{mD}{d} N_{sat} = \frac{mD}{d} \frac{P_{sat}(T_{anode})}{k T_{anode}} \qquad (5)$$

for the planar case. Binary diffusion coefficients, $D_{AB}$, were obtained using the kinetic theory of gases, $D_{AB} = \frac{3\pi}{32}\sqrt{\frac{8kT}{\pi(m_A m_B/m_A+m_B)}}\frac{1}{(n_A+n_B)\sigma_{AB}}$, where A is either iron or carbon atoms, B is background argon gas, and $\sigma_{AB} = \pi(r_A + r_B)^2$ [6,11,31]. In a diffusion-limited regime, the vapor density is negligible compared to the background gas density, i.e., $n_{\text{vapor}} + n_{\text{background gas}} \approx n_{\text{background gas}}$. The saturation vapor pressure of iron was calculated using the expression from ref. [32], and that for graphite using the Clausius–Clapeyron relation (see ref. [6] for details).

Khrabry et al. explained in detail the derivation of the anode ablation flux, starting from the Hertz-Knudsen equation (Eq. 3) [6]. These authors related the vapor partial pressure at the ablation surface, $p_{\text{vapor}}$, to the ablation flux, $g_{\text{abl}}$, and diffusion of the ablated material through the background gas. In the diffusion-limited regime and for the semi-infinite case, they obtained $g_{\text{abl}} = \frac{mnD}{d}\frac{P_{\text{sat}}}{P}$, which is exactly what we derived here for the same case: at the liquid-vapor interface, $g_{\text{abl}} = g_{\text{diff}} = mD\frac{N_{\text{sat}}}{d}$.

The evaporation surface radius, R, was determined for each experiment by drawing a circle around the molten steel anode surface (Figure 3). At high arc current and power (Figure 3a), the anode fully melted and formed a hemisphere, and so the evaporation surface radius could be easily measured. At medium and low arc currents and powers (Figures 3b, 3c, and 3d), the anode melted only partially and formed liquid domes for which the radius can drastically differ from experiment to experiment. For the planar graphite anode, the ablation surface was considered flat, i.e., $R \to \infty$.

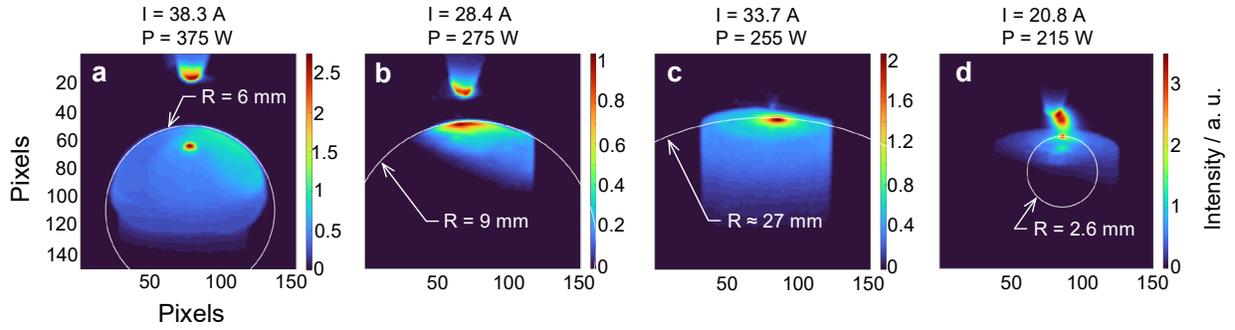

Figure 3. Images of the steel anode at different arc current and power settings. White circles are drawn around the molten anode parts to define the evaporation surface radii. Note that the red spots on the anode surface are the reflections of the hot cathode and do not correspond to high anode temperatures.

To calculate the anode ablation rate, $\Gamma = Ag_{\text{abl}}$, the ablation surface area, A, should be found. For partially molten steel anodes with flat surfaces (Figure 4a, corresponding to Figure 3c), we approximated the anode temperature distribution using circular isothermal areas (shown as white ellipses in Figure 4b). The isotherms appear elliptical due to the angled camera arrangement, which led to the y-axis visually skewed while the x-axis was not distorted. Each annulus area was assigned a single temperature, which was an average between the outer and inner isotherm values (Figure 4c). Thus, for an annulus located between isotherms of 2000 and 2100 K, a temperature of 2050 K was assigned. The areas were calculated using the calibrated camera resolution of 106 µm/pixel or 1 mm = 9.4 pix.

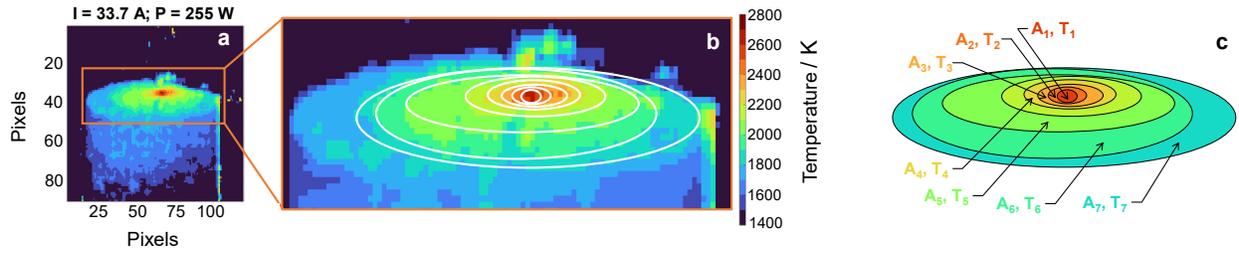

Figure 4. (a) Pyrometric image of a partially molten steel anode; (b) Zoomed-in part of the anode surface with white ellipses denoting isotherms; (c) Schematic showing isothermal areas used to calculate the total anode ablation rate.

The case with a fully molten steel anode is shown in Figure 5. Here, the ablation surface area cannot be assumed flat. Instead, isothermal areas were calculated as $A_i = \Omega_i R^2 = \pi \sin^2\left(\frac{\alpha_i}{2}\right) R^2$, where $\Omega$ is the solid angle encompassing the corresponding isothermal region.

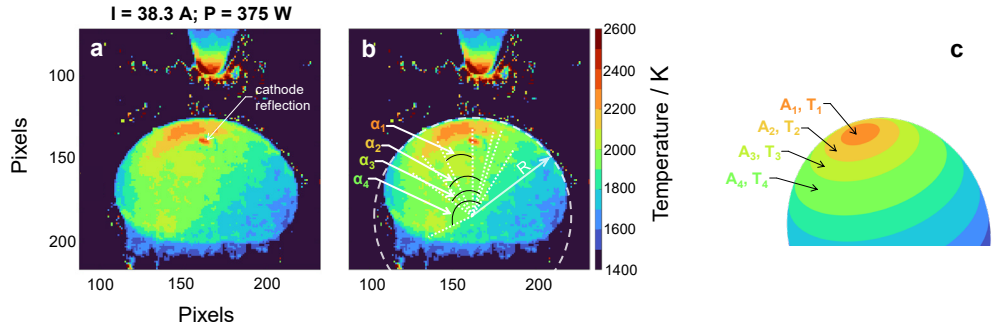

Figure 5. (a) Pyrometric image of a fully molten steel anode; (b) Same image in (a) showing projections of solid angles encompassing the isothermal regions; (c) Schematic showing isothermal areas used to calculate the total anode ablation rate.

The total anode ablation rate is found by summing up the ablation rates corresponding to each isothermal area,

$$\Gamma = \sum_i A_i g_{\text{abl},i} = \sum_i A_i m D_i \frac{R+d}{Rd} \frac{P_{\text{sat}}(T_{\text{anode},i})}{kT_{\text{anode},i}}. \tag{6}$$

The condensation distance, *d*, can be approximated as the interelectrode gap in semi-infinite configurations, where a large flat cathode is located close to a smaller flat anode, e.g., as in the case with graphite electrodes herein and in ref. [6]. The hypothesis is that most of the ablated anode material diffuses in the direction of the cathode and condenses on its surface, which was confirmed in this work and ref. [30]. Therefore, the interelectrode gap of $d$ = 3 mm was used when calculating the ablation rate of the graphite anode. However, for steel anode experiments, the cathode was conically shaped and the anode surface was not always flat, as demonstrated in Figure 3. In that case, the condensation distance (maximum likelihood estimate) was found using least-squares minimization, $d = \underset{d}{\mathrm{argmin}}\{\sum_j(\Gamma_{\mathrm{exp},j} - \Gamma_{\mathrm{pred},j})^2\}$, where $d$ was subject to the constraints $1 \text{ mm} \leq d \leq 10 \text{ mm}$, $\Gamma_{\mathrm{exp},j}$ is the measured anode mass loss rate for the $j$-th experiment, and $\Gamma_{\mathrm{pred},j}$ is the corresponding ablation rate prediction calculated using Eq. 6. We used the "fminbnd" minimization function in MATLAB®, which utilizes a golden section search and parabolic interpolation. This procedure yielded the condensation distance, $d$ = 4.5 mm, and the coefficient of determination, $R^2$ = 0.77, which suggests that 23 % of the variance in the experimental data could not be explained with the simple diffusion-limited evaporation model used herein.

## 5. Results and discussion

### 5.1. Graphite anode in argon

First, we tested the pyrometry system in a plasma environment using a graphite cathode and anode to avoid complications associated with molten metals, such as specular reflections and uncertain emissivity. Graphite optical properties were approximated as those of a grey body, i.e., $\frac{\varepsilon_{785\,\mathrm{nm}}}{\varepsilon_{890\,\mathrm{nm}}} = 1$ [33]. We used a 9.5-mm diameter graphite cathode and a 6.4-mm diameter graphite anode. The gap between the electrodes was held at 3 mm. The experiments were conducted with argon at 66.7 kPa (500 Torr). Two arc settings were used: 38.4 A, 500 ± 40 W; and 53 A, 750 ± 50 W. The power variation was caused by the unstable arc behavior that led to fluctuating voltages, while the current was fixed by the power supply. Each experiment (including those described in further sections with a steel anode) lasted for 30—60 minutes, during which multiple pyrometric images were taken.

In Figure 6, the top row shows the emission intensity of the graphite anode at λ = 890 nm, and the bottom row shows the corresponding anode temperatures and calculated ablation rates. The actual mass loss rate was measured to be 97 mg/h. Here and in the rest of the manuscript, a mean 3x3 spatial filter was applied to reduce the pixel noise in figures showing temperature.

The arc discharge between two graphite electrodes is notoriously unstable with a "wandering" arc attachment spot [15]. At $t_0$, the anode attachment spot was facing the camera, resulting in a high-intensity emission signal (Figure 6a), temperatures up to 3450 K (Figure 6g), and a calculated ablation rate of 163 mg/h. In Figure 6b, the arc attachment spot moved out of view due to an incoming shutter (not distinguishable in the figure), resulting in a lower signal intensity. The anode temperature dropped by ~200 K (Figure 6h), leading to a significantly reduced ablation rate estimation of 77 mg/h. A temperature drop of 200 K within 0.5 ms cannot be explained by ablation, radiation, or conduction heat losses. For the remaining timeframe depicted in Figure 6, the arc attachment stayed outside of the camera field of view. Anode temperatures and calculated ablation rates remained stable for 1—2 ms, noticeably reducing only after 5 ms. Abrupt temperature changes were also observed in experiments with a steel anode when the arc attachment moved out of view, e.g., the surface temperature dropped by ~200 K in 0.5 ms and then stabilized for 1—2 ms (experiments at 2000 fps) or the surface temperature dropped by ~600 K in 1.67 ms (experiments at 600 fps). We conclude that pyrometric imaging of the arc attachment region yields erroneous temperatures and should be avoided. A possible explanation could be local interfering emission, such as from plasma radiation scattered by vapor particles.

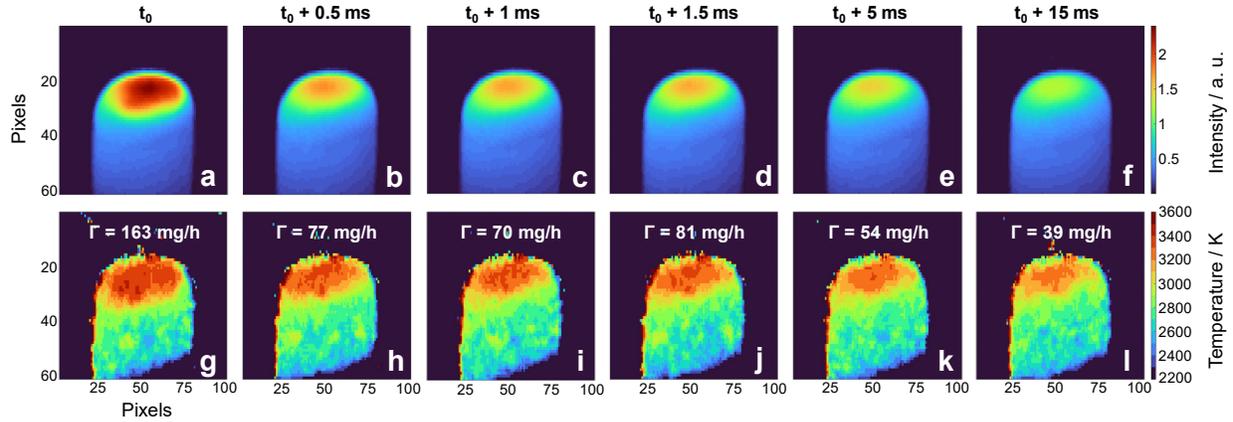

Figure 6. The effect of the arc attachment spot on the graphite anode temperature and calculated ablation rate. Top row: Emission intensities at λ = 890 nm captured at different time delays. Bottom row: Corresponding pyrometric images and the calculated ablation rates. The experimentally measured anode mass loss rate was 97 mg/h. Arc discharge conditions: I = 53 A, P = 750 ± 50 W. At $t_0$, the arc attachment spot is in the camera field of view, resulting in high emission intensity, temperature, and calculated ablation rate. For the rest of the depicted timeframe, the arc attachment is out of the field of view. It is suggested that pyrometric imaging of the arc attachment region yields erroneous temperatures and should be avoided.

Figures 7a and 7b show typical pyrometric images of the graphite anode obtained at different arc settings (taken when the anode attachment spot was outside of the camera's field of view) and the estimations of the corresponding ablation rates. Figure 7c compares the calculated ablation rates versus measured anode mass loss rates. Vertical error bars denote a ± 60% uncertainty that originates from a ± 3.5% uncertainty in pyrometric temperatures. A reasonable agreement between the predicted and measured ablation rates supports the reliability of our pyrometric imaging and the chosen ablation rate model.

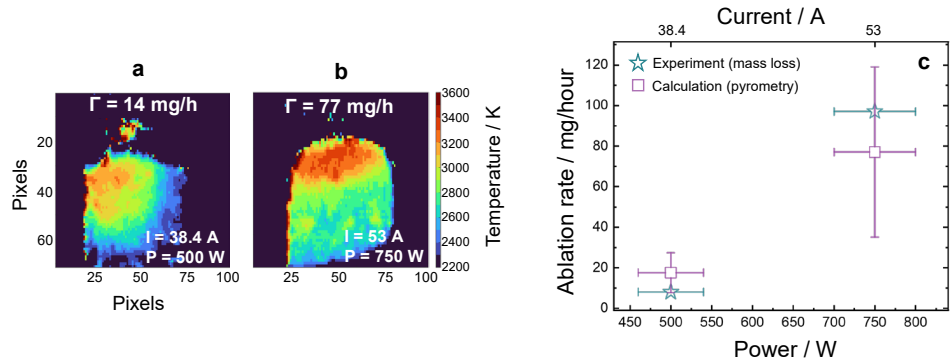

Figure 7. (a,b) Typical pyrometric images of a graphite anode at different arc settings; (c) Calculated and measured graphite anode ablation rates. Vertical error bars denote a ± 60% uncertainty that originates from a ± 3.5% uncertainty in the pyrometric temperatures.

## 5.2. Steel anode in argon

Unlike graphite, the surface of a molten metal acts as a mirror that reflects external radiation, which can lead to erroneous temperature measurements. This is highlighted in Figure 8, where the top row of images shows thermal emission intensities at λ = 890 nm, and the bottom row of images shows corresponding pyrometric temperatures. Throughout the depicted sequence, reflections of the hot cathode and dense plasma plume can be seen on the anode surface. These reflections resulted in drastically different temperatures compared to those of the surrounding areas, often yielding non-physical values, e.g., exceeding the boiling point of iron. At $t_0$ + 5 ms (Figures 8f and 8n), the incoming shutter extinguishes the arc, eliminating the reflection of the plasma plume. At $t_0$ + 7 ms (Figures 8h and 8p), the shutter continues its swing and blocks the cathode reflection, which enables uncorrupted measurements of the anode surface temperature. A high-speed video corresponding to Figure 8 that shows the swing of the shutter is available in the Supplementary Information. Although the shutter disturbs the plasma and can sometimes lead to plasma extinguishment, it enables the anode temperature measurements. We note that in this experiment, the lower left side of the camera field of view was obstructed by an optical flange edge.

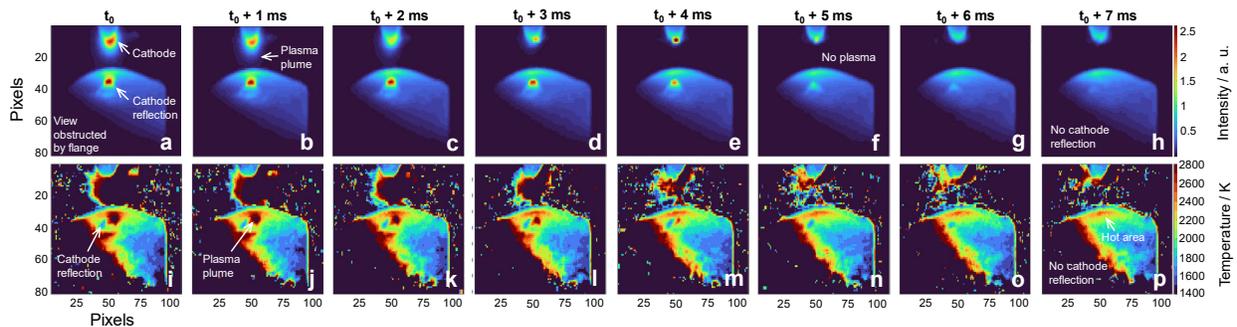

Figure 8. Effect of reflections on pyrometric temperature measurements. Top row: Emission intensities of a steel

anode at λ = 890 nm captured at different time delays. Bottom row: Corresponding pyrometric images for those images located directly above. At $t_0$ + 5 ms, the incoming shutter blocks the reflection from the dense plasma plume. At $t_0$ + 7 ms, the shutter blocks the reflection from the hot cathode and enables an uncorrupted anode temperature measurement. The lower left side of the field of view was obstructed by the edge of an optical flange. Arc discharge conditions: I = 28.8 A, P = 270 ± 10 W.

Figure 9 illustrates the dynamic behavior of the liquid metal surface, which we attribute to the Marangoni effect [17]. Localized arc heating of the liquid anode creates a temperature gradient, inducing thermocapillary forces that drive liquid motion due to surface tension variations. The colorbar scale in Figure 9 was intentionally constrained to highlight the liquid metal movement and the associated changes in surface temperature. The depicted time sequence shows how the hot liquid metal (depicted in green in the top row, corresponding to T = 2100—2200 K in the bottom row) moves from the right side of the anode to its center and then spreads down as it cools. The measured liquid velocity ranged between 4—10 cm/s, and the typical timescale, at which spatial temperature variations were distinguishable, was a few milliseconds, which necessitated the use of high-speed pyrometry. The video corresponding to Figure 9 is available in the Supplementary Information.

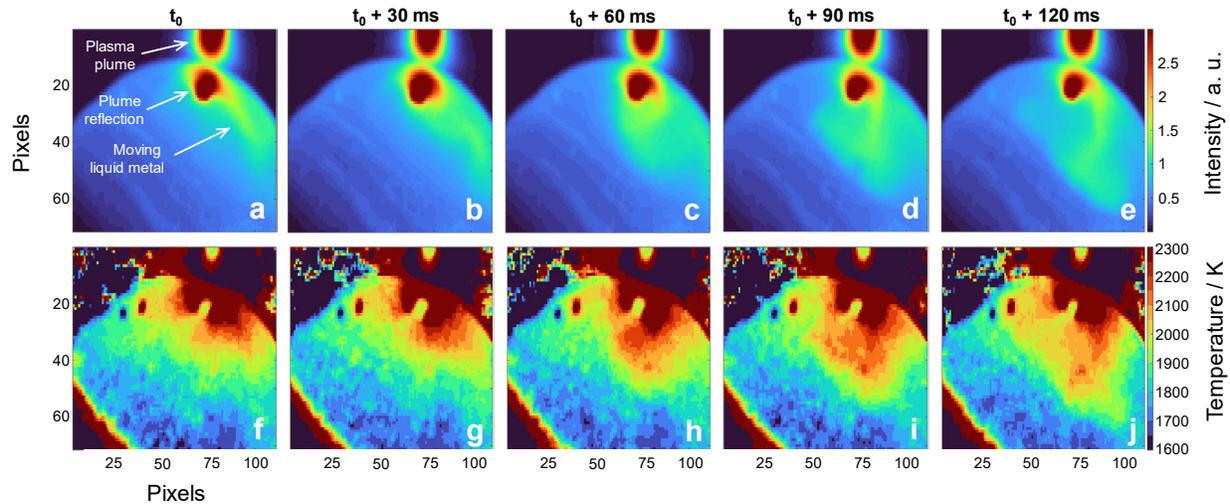

Figure 9. The effect of liquid metal motion (Marangoni effect) on the surface temperature distribution. Top row: Emission intensities of a steel anode at λ = 890 nm captured at different time delays. Bottom row: Corresponding pyrometric images for those images located directly above. The hot liquid metal (depicted in green in the top row, corresponding to T = 2100—2200 K in the bottom row) moves from the right side of the anode to its center and then spreads downward as it cools. Arc discharge conditions: I = 48 A, P = 465 ± 15 W.

To support the attribution of the liquid motion to the Marangoni effect, we calculated the liquid velocity following the approach from ref. [34]. For large Reynolds numbers (Re), the liquid motion takes place primarily in a thin "boundary layer" [35]. In that case, the liquid velocity at the surface can be estimated as $v \sim \left(\frac{R}{\nu}\right)^{1/3} \left(\frac{1}{\rho}\left|\frac{d\sigma}{dT}\right|\frac{\Delta T}{L}\right)^{2/3}$, where $R \sim 1$ mm is the size of the heat source (assumed to be the hottest spot on the anode surface), $\nu = 5 \cdot 10^{-7}$ m$^2$/s is the steel kinematic viscocity, $\rho = 7000$ kg/m$^3$ is the liquid steel density, $\frac{d\sigma}{dT} = -3.5 \cdot 10^{-4}$ N m$^{-1}$ K$^{-1}$ is the temperature-dependent surface tension, and $\Delta T = 200$ K is the experimentally measured temperature change over the distance $L \sim 5$ mm. All material properties were taken from ref. [34]. The calculation yields velocities between 10—20 cm/s, which is of the same order of magnitude as the experimentally measured velocities, supporting our assertion that the liquid movement is due to the Marangoni effect. The assumption of a large Re number is justified by the calculation of the Re value, $Re = \frac{Lv}{\nu} \approx 1000$.

Figure 10 shows typical pyrometric images of a steel anode in an arc in an argon atmosphere. We note the different ranges of the colorbars in Figures 10a-c. At low current (< 25 A) and power (< 230 W) (Figure 10a), the anode primarily remained solid with a small molten region on its top. The surface reached temperatures up to 2300—2400 K within a small spot of 1—2 mm, resulting in predicted ablation rates under 5 mg/h. At an arc current between 25—35 A and power 230—300 W (Figure 10b), the top surface of the anode significantly melted and formed a shallow puddle. The highest observed temperatures at these arc settings were between 2400—2600 K, with puddle dimensions significantly varying among the experiments, which led to a spread of predicted ablation rates between 7—27 mg/h. Further increases in the arc current, and hence power, led to extensive anode melting and its reshaping into a hemisphere (Figure 10c). Despite a high arc power, the maximum temperature did not exceed 2300—2400 K. We suggest that this is due to efficient thermal dissipation by the Marangoni effect within the molten pool (see Figure 9). Nevertheless, large ablating surface areas led to higher predicted ablation rates ranging between 25—33 mg/h in this regime.

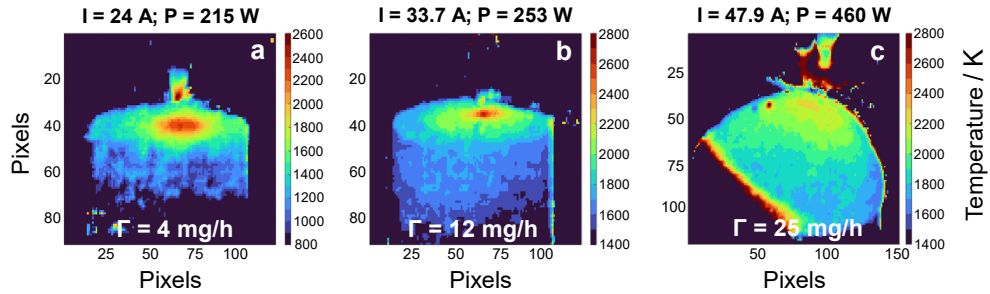

Figure 10. Typical pyrometric images of a steel anode in argon at different arc settings. Note the increasing dimensions of the molten metal pool (areas corresponding to temperatures > 1800 K) with increasing arc current and power. Ablation rates, $\Gamma$ (mg/h), were calculated using the pyrometric temperatures.

Predicted and measured ablation rates of the steel anode in an arc discharge in argon are shown in Figure 11. While the arc current was fixed during an experiment, the voltage could fluctuate due to the arc movement and slight changes of the interelectrode gap (e.g., melting of the anode). To assess this possible effect, Figure 11b shows the dependence of ablation rates on arc power, $P = I \times U$. Unlike that for a planar graphite anode, calculations for this case involved an adjustable parameter, i.e., the condensation distance, $d$ = 4.5 mm, which was determined using least-squares minimization, as described in Section 4. Considering the use of an adjustable parameter and the uncertainties in pyrometric temperatures, iron saturation vapor pressure, and diffusion coefficients, we cannot expect to report an exact match with the experiments. Nonetheless, the identified trends should hold regardless of the chosen condensation distance. We point out that no adjustable parameters were used to achieve the quantitative agreement between the predicted and measured graphite ablation rates (Figure 7c), which corroborates the reliability of the anode temperature measurements and supports our selection of the ablation model.

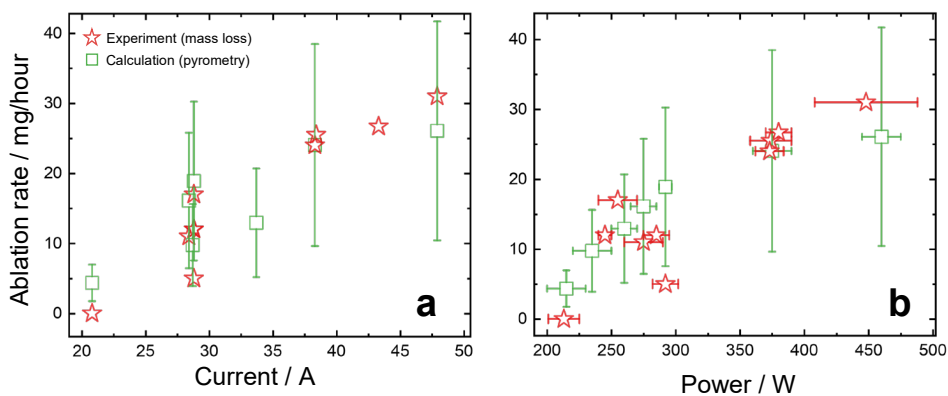

Figure 11. Calculated and experimentally measured ablation rates of a steel anode in an arc in argon. Vertical error bars denote a ± 60% uncertainty that originates from a ± 3.5% uncertainty in pyrometric temperatures. Calculations of the steel anode ablation rates involved an adjustable parameter, the condensation distance, $d$ = 4.5 mm, which was determined by using least-squares minimization, yielding $R^2$ = 0.77

### 5.3. Steel anode in Ar/CH$_4$

Finally, we conducted experiments in an argon/methane atmosphere (Ar 97.6 wt.%, CH$_4$ 2.4 wt.%) under conditions relevant for SWCNT synthesis [5]. Figure 12 shows the image of the steel anode emission intensity and the corresponding pyrometric image. We note that with CH$_4$ addition, higher arc voltages were required to maintain the same arc current compared to those in pure Ar. This is because of the energy consumption for methane dissociation and the higher thermal conductivity of CH$_4$ and its decomposition products, such as H$_2$, compared to Ar. This leads to arc cooling and hence, its increasing resistivity. For a current of 28.8 A, the resulting arc power was 430 ± 10 W, which is significantly higher than 235—300 W in pure Ar for the same current.

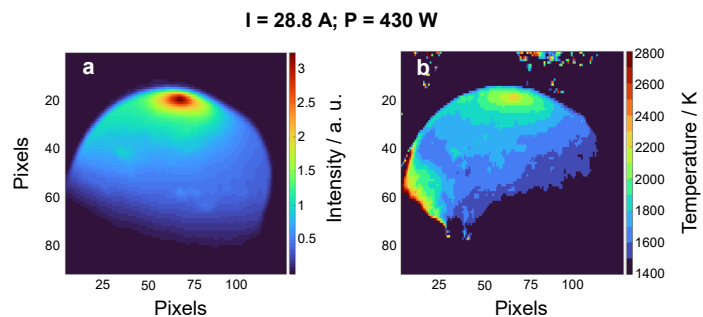

Figure 12. Images of a steel anode in an argon/methane atmosphere at an arc current of 28.8 A and arc power of 430 ± 10 W. (a) Emission intensity of the steel anode captured at λ = 890 nm. (b) The corresponding pyrometric image.

It was not possible to directly measure the anode mass loss rate because of carburization, i.e., diffusion of carbon into the molten steel. Indeed, the steel anode during operation gained mass at a typical rate of 100—200 mg/h. Therefore, the anode ablation rate, and hence the production rate of metal catalyst nanoparticles, a crucial parameter for SWNCT synthesis, could only be predicted using anode temperature imaging. As shown in Figure 12, the anode surface was fully molten with a maximum temperature between 2200—2300 K. This is comparable to the maximum temperature measured in argon at a similar arc power (e.g., Figure 10c). However, the high temperature regions were much more constricted than those observed in pure argon (Figure 12b vs. Figure 10c).

Evidently, carburization of the steel anode changes its temperature profile. First, with increasing carbon content, the steel melting point decreases, leading to a larger molten metal pool and hence, larger area of heat dissipation by Marangoni waves. Second, with increasing carbon concentration, the surface tension of the molten iron decreases [36], which should lead to enhanced Marangoni waves and enhanced heat dissipation, explaining the reduced high temperature regions. Once cooled, carbon should precipitate on the anode surface. Indeed, we confirmed high carbon concentrations at the anode surface using X-ray photoelectron spectroscopy (XPS), as described in the Supplementary Information. Furthermore, the energy-dispersive X-ray spectroscopy (EDS) analysis of the anode cross-section revealed that carbon had diffused tens of micrometers into the anode (Figure S7 and Table S3).

The anode ablation rate was calculated assuming the same condensation distance as in pure argon experiments: $d$ = 4.5 mm. This assumption was based on our previous modeling work, which investigated iron particle growth at the same conditions, namely in Ar and in Ar + 2.4 wt.% $CH_4$ [10]. In that study, we achieved good agreement between modeled and measured iron particle sizes across both gas environments, without modifying the diffusion coefficient, which defines the condensation distance. Therefore, the condensation distance of $d$ = 4.5 mm was used, resulting in the anode ablation rate of 2—8 mg/h, which is notably lower than the 10—40 mg/h ablation rate predicted in pure argon for a similar arc power (Figure 11 b), and 4—30 mg/h ablation rate in pure argon for the same arc current (Figure 11 a).

Importantly, the predicted reduction of the anode ablation rate in a hydrocarbon environment translates into smaller production rate of metal catalyst nanoparticles. This effect must be considered when scaling up the production of SWCNTs, e.g., when optimizing for the catalyst-to-carbon ratio and for the catalyst particle size [5,10].

# 6. Conclusions

We have investigated the ablation of a molten steel anode in a DC arc with methane, in conditions relevant for the production of single-walled carbon nanotubes (SWCNTs). Measuring the anode ablation by weight before and after synthesis was not feasible due to the anode carburization that occurs in the hydrocarbon atmosphere. To overcome this issue, we implemented a high-speed, 2D, 2-color pyrometry for in situ temperature measurements of the molten steel anode. The obtained temperature fields were used to calculate the steel anode ablation rate using a diffusion-limited evaporation model. In experiments conducted under an argon atmosphere, i.e., without anode carburization, the predicted ablation rates agreed with the experimentally measured anode mass loss rates. Substantially lower ablation rates were predicted for steel anodes exposed to methane compared to those exposed to pure argon, despite the use of similar arc power in the two cases. This effect must be considered when scaling up the production of SWCNTs, e.g., when optimizing for the catalyst-to-carbon ratio. In addition, we found that the Marangoni effect could significantly influence heat dissipation in the molten anode, potentially limiting its ablation rate.

## Acknowledgments

This work was supported by the U.S. Department of Energy through contract DE-AC02-09CH11466. We thank Ivan Romadanov and Nirbhav Chopra (both PPPL) for fruitful discussions, Timothy Bennett and Alexander Merzhevsky (both PPPL) for engineering support, Adam McLean and Brentley Stratton lending the calibration light source, and Kaifu Gan (all PPPL) for lending the black body source.